\documentclass{pasj00}

\def\kms{~km~s$^{-1}$}
\def\h2o{H$_2$O}
\def\vlsr{$V_{\mbox{\scriptsize LSR}}$\ }
\def\etal{ et~al.\ }
\def\i2248{IRAS~22480$+$6002}
\def\j2254{J225425.3$+$620939}

\SetRunningHead{H.~Imai\etal}{\h2o masers in \i2248}

\title{Annual Parallax Distance to the K-type star system \i2248 measured with VERA}
\author{Hiroshi \textsc{Imai}\altaffilmark{1,2},
Nobuyuki \textsc{Sakai}\altaffilmark{3}, 
Hiroyuki \textsc{Nakanishi}\altaffilmark{1}, \\ Hirofumi \textsc{Sakanoue}\altaffilmark{1}, 
Mareki \textsc{Honma}\altaffilmark{4}, and Takeshi \textsc{Miyaji}\altaffilmark{4}}
\altaffiltext{1}{Department of Physics, Graduate School of Science and Engineering, \\
Kagoshima University, Kagoshima 890-0065}
\altaffiltext{2}{International Centre for Radio Astronomy Research, M468, \\
The University of Western Australia, 35 Stirling Hwy, Crawley, Western Australia, 6009} 
\altaffiltext{3}{Department of Astronomical Science, Graduate University for Advanced Studies, 
Mitaka, Tokyo 181--8588}
\altaffiltext{4}{Mizusawa VLBI Observatory, National Astronomical Observatory, 
Mitaka, Tokyo 181--8588}
\email{(HI) hiroimai@sci.kagoshima-u.ac.jp}
\Received{2012/04/28}
\Accepted{2012/08/02}
\Published{2013/02/25}

\begin{document}

\maketitle

\begin{abstract}
We present the astrometric VLBI observations of \h2o masers associated with  \i2248 ($=$ IRC$+$60370, hereafter I22480) with the VLBI Exploration of Radio Astrometry (VERA). The stellar type of I22480 looks unusual as a stellar maser source and has been debated since the 1970s. We successfully determined the annual parallax of a group of the \h2o maser spots,  $\pi=0.400\pm 0.025$~mas, corresponding to a distance to I22480 of $D=2.50^{+0.17}_{-0.15}$~kpc. This suggests that the estimated bolometric luminosity of I22480 should be revised to 35 000 $L_{\odot}$, favoring a K-type supergiant rather than an RV Tau-type variable star previously suggested. Although the spectral type is unusual as a stellar maser source, the internal motions of the \h2o maser features suggest that the \h2o masers are associated with the circumstellar envelope of this star. Taking into account a possible stellar motion with respect to the maser feature motions, we derived a secular proper motion of I22480, $(\mu_{\alpha}, \mu_{\delta})=(-2.58\pm 0.33, -1.91\pm 0.17)$ [mas~yr$^{-1}$]. The derived motion of I22480 in the Milky Way has a deviation by $\sim -$30\kms\ in the galactic azimuthal direction from a circular motion estimated from the galactocentric distance to I22480 and assumption of a flat Galactic rotation curve. This peculiar motion is still comparable to those typically seen in the \h2o\ maser sources located in the Perseus spiral arm. Taking into account the peculiar motion and the proximity to the Galactic midplane ($z\simeq$60~pc), I22480 may be a member of the Galactic thin disk. 
\end{abstract} 

\section{Introduction}
\label{sec:introduction}
\h2o maser emission is observed in circumstellar envelopes of evolved stars such as Mira variables, OH/IR stars, and red supergiants, which exhibit energetic stellar mass loss in a rate of $\dot{M} \geq 10^{-7}M_{\odot}$~yr$^{-1}$ \citep{eli92}. Most of these stars with \h2o (and SiO) maser emission have the spectral type M, except central stars of pre-planetary and planetary nebulae, whose circumstellar envelopes are destroyed (e.g., \cite{mir01}), or stars whose surfaces are still too hot but that will become M-type stars. 

\i2248  ($=$AFGL~2968, IRC$+$60370, hereafter abbreviated as I22480) has been classified as a K-type supergiant \citep{hum74,faw77}. Although this star has a $V$ magnitude of $\sim$8.3 (Tycho Input Catalogue), 
it has not been reported as a variable star. The detection of \h2o and SiO masers was surprising for the reason 
mentioned above \citep{han98,nym98}. However, \citet{win94} gave a new spectral classification of M01 for I22480. CO emission was detected towards I22480, indicating that energetic mass loss still/already occurs. The systemic stellar velocity and the mass-loss flow velocity of I22480 are estimated to be \vlsr$=-$49.3\kms\ and 26.4\kms, respectively \citep{gro99}, which are consistent with those estimated from the spatio-kinematics of the \h2o masers in I22480 \citep{ima08}. The Two-Micron Sky Survey (2MASS) image shows a secondary star that is located only $\sim$12\arcsec\ east of I22480, but this B5II star may not have a physical association of the western K-type star. The distance to I22480 was estimated to be 5.0~kpc on the basis of the kinematic distance method. This gives the K-type star an extremely high luminosity $L_{\ast}=$140 000~$L_{\odot}$ \citep{gro99}. On the contrary, \citet{ima08} obtained a distance value of $\sim$1~kpc on the basis of the statistical parallax method applied for the \h2o\ masers in I22480. If it is true, the stellar luminosity should be down to  $L_{\ast}=$5 800~$L_{\odot}$, favoring an RV Tau-type variable star, or a population II low-mass post-AGB star. 

In this paper, we report new VLBI astrometric observations of \h2o\ masers associated with I22480 with 
the VLBI Exploration of Radio Astrometry (VERA)\footnote
{Mizusawa VERA observatory is a branch of the National Astronomical Observatory, an 
interuniversity research institute operated by the Ministry of Education, Culture, Sports, Science and 
Technology.}, 
which yielded successful detection of the annual parallax of I22480. 

%%%%%%%%%%% Put here Tables 1 %%%%%%%%%%%%%

\section{Observations and data reduction}
\label{sec:observations}
The VERA observations of the I22480 \h2o\ ($J_{K_{-}K_{+}}=6_{12}-5_{23}$, 22.235080~GHz) masers were conducted at 16 epochs from 2009 December to 2011 December. Table \ref{tab:status} gives a summary of these observations and radio source mapping. At each epoch, the observation was made for 8 hr in total. I22480 was scanned for $\sim$4.5~hr together with the position reference source, \j2254 (hereafter abbreviated as J2254), simultaneously using VERA's dual-beam system. I22480 and J2254 are separated by 1\arcdeg.94. 3C454.3 was also scanned in both of the dual beams for 5~min per beam every $\sim$40~min for calibration of group delay residuals. The received signals were digitized in four quantization levels, then divided into 16 base band channels (BBCs) with a bandwidth of 16 MHz each. The signals in one of the BBCs were obtained from scans on I22480 at the frequency band including the \h2o\ maser emission, while those in other BBCs from scans on J2254 with a total frequency band range of 480~MHz. The BBC outputs were recorded in a rate of 1024~Mbits~s$^{-1}$.  A data correlation was made with the Mitaka FX correlator. The accumulation period of the correlation was set to 1~s. The correlation outputs consisted of 512 and 32 spectral channels for the \h2o\ maser and continuum emission, respectively. The former corresponds to a velocity spacing of 0.42\kms, which is narrow enough to resolve an \h2o {\it maser feature} (corresponding to a maser gas clump), which consists of two or more spectral channel components called {\it maser spots}. 

Data reduction was mainly made with the National Radio Astronomy Observatory (NRAO) Astronomical Image Processing System (AIPS) package. For astrometry, we need special procedures described as follows (see also e.g., \cite{hon07,ima07}). At first, delay-tracking was again performed for the correlated data using better delay-tracking models calculated with the original software equivalent to the CALC9 package developed by the Goddard Space Flight Center/NASA VLBI group. Through the whole data analysis, we adopted the coordinates of the delay-tracking center: 
$\alpha_{J2000}=$22$^{\mbox h}$49$^{\mbox m}$58$^{\mbox s}$\hspace{-2pt}.876,  
$\delta_{J2000}=$~$+$60$^{\circ}$17$^{\prime}$56\arcsec.65 for I22480 and 
$\alpha_{J2000}=$22$^{\mbox h}$54$^{\mbox m}$25$^{\mbox s}$\hspace{-2pt}.293179,  
$\delta_{J2000}=$~$+$62$^{\circ}$09$^{\prime}$38\arcsec.72393 for J2254. 
The delay-tracking solutions include residual delay contributions from the atmosphere, which were estimated using the global positioning system (GPS) data \citep{hon08b}. Secondly, differences in instrumental delays between two signal paths in the dual beam system were calibrated using the differential delays, which were measured using artificial noise signals injected into the two receivers during the observations \citep{hon08a}. Because J2254 was too faint ($\sim$12~mJy) to detect in the normal fringe-fitting procedure, this should be detected through {\it inverse} phase-referencing to a bright \h2o maser spot in I22480 as mantioned soon later. Thirdly, fringe-fitting was performed using the data of 3C454.3, whose solutions of group delay, rate (fringe frequency), and phase residuals for  clock parameter calibration were applied to the data calibration of both I22480 and J2254. Fourthly, fringe-fitting and self-calibration were performed using the brightest \h2o maser spot in I22480, whose velocity with respect to the local standard of rest (LSR) is given 
in Column 8 of table \ref{tab:status}, without solving delay residuals, in a solution interval of 2~min or shorter,  whose solutions were applied to the data calibration of both I22480 and J2254. Before applying the second fringe-fitting solutions to the data calibration of J2254, the data of J2254 were averaged into a single spectral channel. The sequence of AIPS data reduction in the inverse phase-referencing part is described in Appendix \ref{sec:appendix1} in more detail. Finally, image cubes of the maser source and the image of J2254 were obtained in visibility deconvolution through the CLEAN algorithm in a typical synthesized beam of 1.3$\times$0.9 in milliarcseconds (mas). Each maser spot (or velocity component) was automatically identified as a Gaussian brightness component using the AIPS task SAD. The parameters of J2254 as a Gaussian component was obtained using AIPS task JMFIT. 

\section{Results}
\label{sec:results}

%%%%%%%%%%% Put here Tables 2, 3 %%%%%%%%%%%%%
%%%%%%%%%%% Put here Figures 1, 2 %%%%%%%%%%%%%

\subsection{The spatio-kinematics of \h2o\ masers in \i2248}
\label{sec:spatio-kinematics}

\begin{figure}[p]
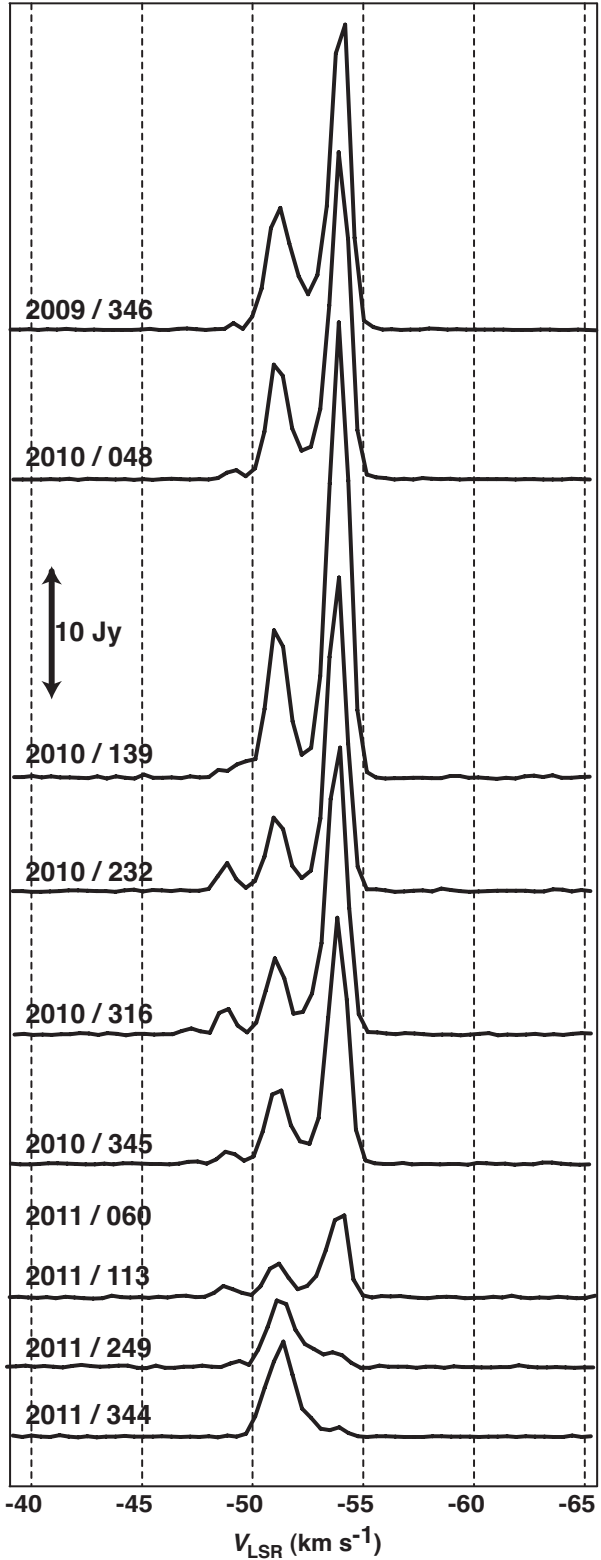

  \begin{center}
     \FigureFile(0.9\hsize,\hsize){fig1.eps}
  \end{center}
  \caption{Scalar-averaged cross-power spectra of \i2248, which were synthesized 
  with the data from all baselines. For clarity, these spectra are shifted in the vertical 
  direction along with the observed epochs. The dates of observation epochs 
  (year / day of the year) are also displayed. }
\label{fig:I2248_spectrum}
\end{figure}

Figure \ref{fig:I2248_spectrum} shows the variation of the cross-power spectrum of the I22480 \h2o masers. The \h2o masers covered a velocity range of $-40$\kms$<$\vlsr$<-60$\kms; such a velocity width is typically seen in Mira-type AGB stars (e.g., \cite{tak94}). \citet{ima08} found five spectral peaks with roughly equal separations of 2--3\kms\ between the peaks. In the present spectra, only middle three out of the five peaks at \vlsr$\sim -54$, $-52$, and $-48$\kms\ were clearly seen. The $-53.9$\kms\ (or $-54.1$\kms) component (maser spot) in the $-$54\kms\ peak was always visible through the observation epochs and suitable for visibility phase calibration and as position reference for the VERA astrometry. In the latest two epochs when this component weakend, its position was determined with respect to the $-51.0$\kms\ (or $-51.4$\kms) component that was used for the calibration. Although the phase-referencing velocity channel changed in the different observing epochs, as shown in Table \ref{tab:status}, the positions of all maser spots in I22480 as well as that of J2254 were measured with respect to the phase center that was set after fringe-fitting and self-calibration using a specific brighter maser component in the phase-referencing velocity channel (Column 8 of Table \ref{tab:status}). 

\begin{figure}[p]
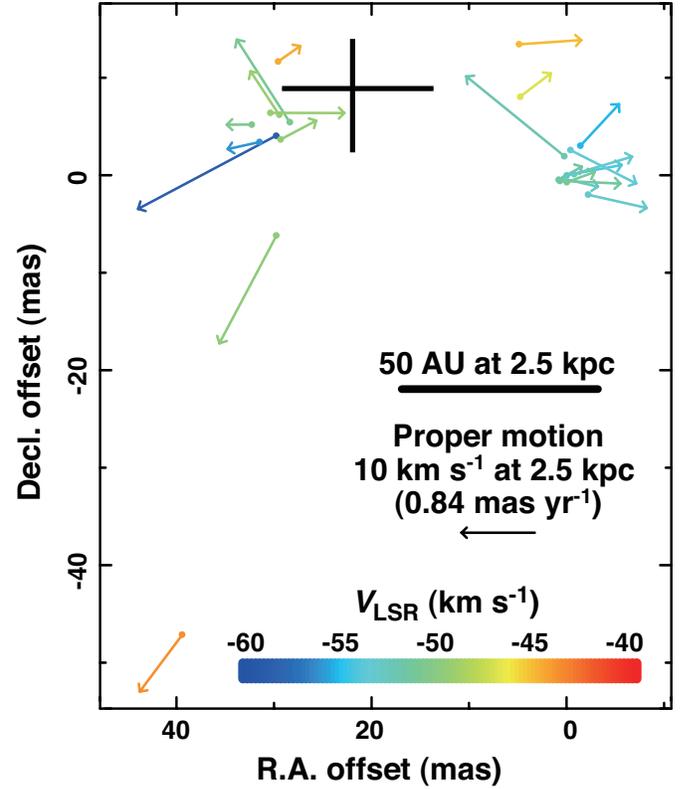

  \begin{center}
    \FigureFile(\hsize,\hsize){fig2.eps}
  \end{center}
\caption{Distribution of \h2o maser features in \i2248. The origin of coordinates is set to the maser 
feature that includes the $-53.9$\kms\ component in the maser feature \i2248:I2013-{\it 4}. 
Colors of maser feature indicate LSR velocities. An arrow shows the relative proper motion of 
the maser feature. The root position of an arrow indicates the location of the maser feature at the 
first of the epochs when the feature was detected. The length and the direction of an arrow indicate 
the speed and direction of the maser proper motion, respectively. The mean proper motion 
$(\dot{X},\dot{Y})=(7,-2)$ [km~s$^{-1}$] is subtracted from the individual proper motions. 
A plus sign indicates the location of the originating point of the outflow or the star itself, which is 
estimated in the model fitting using the maser kinematical data. The size of the symbol indicates 
the position error.}
\label{fig:proper-motions}
\end{figure}

Figure \ref{fig:proper-motions} shows the distribution of \h2o maser features and their relative proper motions. Table \ref{tab:pmotions} gives the parameters of the maser features whose proper motions were identified. The angular extent of the maser distribution in the present work was up to $\sim$70~mas and larger than that previously reported \citep{ima08}. The difference in the extent is attributed to a new maser feature found in the southeast side of the maser distribution. Two major clusters of maser features were always seen in the east--west direction with a separation of $\sim$35~mas. One can roughly see outflow motions of maser features from a common originating point (the possible position of the central star). This may rule out the possibility previously suggested that the maser feature alignment in the north--south direction is attributed to the interaction of the circumstellar envelope harboring the \h2o masers with the wind from the eastern star \citep{ima08}. 

In order to derive the kinematical parameters of the outflow, we performed the least-squares method for the model-fitting analysis as presented by \citet{ima11a}. Throughout the model fitting, we derive a position vector of the originating point of outflow in the maser map, $(\Delta X_{\rm 0}, \Delta Y_{\rm 0})$ and a velocity vector of the originating point, $(V_{0X}, V_{0Y})$. They were estimated by minimizing a $\chi^2$ value, 

\begin{eqnarray}
\nonumber
\chi^{2}& = & \frac{1}{3N_{\rm m}-N_{\rm p}}
\sum^{N_{\rm m}}_{i}
\left\{
\frac{\left[\mu_{ix}-w_{ix}/(a_{{\rm 0}}D)\right]^{2}}{\sigma^{2}_{\mu_{ix}}} \right. \\
& & 
\left. +\frac{\left[\mu_{iy}-w_{iy}/(a_{{\rm 0}}D)\right]^{2}}{\sigma^{2}_{\mu_{iy}}}
+\frac{\left[u_{iz}-w_{iz}\right]^{2}}{\sigma^{2}_{u_{iz}}} 
\right\}. 
\label{eq:model-fit}
\end{eqnarray}

\noindent
Here $N_{\rm m}$ is the number of maser features with measured proper motions, $N_{\rm p}$ the number of free parameters in the model fitting, $a_{\rm 0}=$4.74\kms~mas$^{-1}$yr~kpc$^{-1}$ a conversion factor from a proper motion to a transverse velocity, and $D\equiv$2.5~kpc the distance to the maser source from the Sun (see Section \ref{sec:annual-parallax}), respectively. $\mu_{ix}$ and $\mu_{iy}$ are the observed proper motion components in the R.A. and decl. directions, respectively, $\sigma_{\mu_{ix}}$ and $\sigma_{\mu_{iy}}$ their uncertainties, $u_{iz}$ the observed line-of-sight(LSR) velocity, and $\sigma_{iz}$ its uncertainty. For simplicity we assume a spherically expanding outflow. In this case, the modeled velocity vector, {\boldmath{$w_{i}$}} $(w_{ix},w_{iy},w_{iz})$, is given as 

\begin{equation}
\mbox{\boldmath $w_{i}$} =  
\mbox{\boldmath $V_{\rm 0}$}(V_{{\rm 0}X},V_{{\rm 0}Y}, V_{{\rm 0}Z})+
V_{\rm exp}(i)\frac{\mbox{\boldmath $r_{i}$}}{r_{i}}, \\
\label{eq:wi}
\end{equation}

\noindent 
where
\begin{eqnarray}
\mbox{\boldmath $r_{i}$} & = & \mbox{\boldmath $x_{i}$}(x_{i}, y_{i}, z_{i})
-\mbox{\boldmath $x_{\rm 0}$}(\Delta X_{\rm 0}, \Delta Y_{\rm 0}), \\
\label{eq:ri}
z_{i} & = & \frac{(u_{iz}-V_{{\rm 0}z})(r^{2}_{ix}+r^{2}_{iy})}
{(u_{ix}-V_{{\rm 0}x})r_{ix}+(u_{iy}-V_{{\rm 0}y})r_{iy}}, \\
\label{eq:zi}
u_{ix} & = & \mu_{ix}a_{\rm 0}D,\;\; u_{iy}=\mu_{iy}a_{\rm 0}D.
\label{eq:uix-uiy}
\end{eqnarray}

\begin{figure}[p]
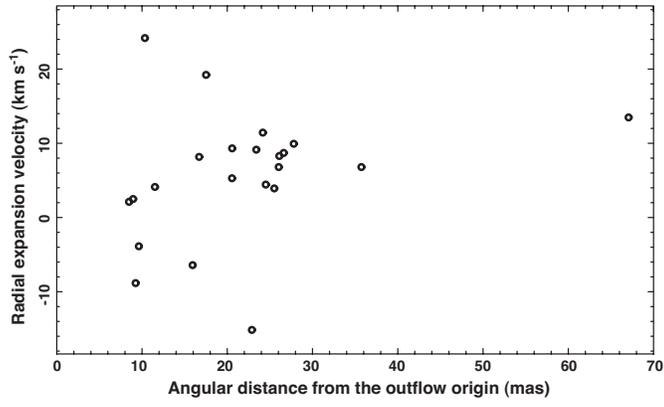

  \begin{center}
       \FigureFile(\hsize,\hsize){fig3.eps}
  \end{center}
  \caption{Distribution of the expansion velocities of the individual
maser features, which are derived from the model fitting.}
\label{fig:expansion-velocity}
\end{figure}

\noindent
The systemic LSR velocity of the central star, $V_{{\rm 0}Z}=-50$\kms, is adopted in the fitting. Table \ref{tab:kinematic-model} gives the derived parameters of the fitting. A plus sign in figure \ref{fig:proper-motions} indicates the estimated location of the originating point of the outflow. The location looks biased to the eastern maser feature cluster due to relatively high velocity components of the maser motions. We do not discuss the location of the central star in this paper. The radial expansion velocities of the individual features are calculated after the model fitting (eq.\ 7 of \cite{ima11a}). Figure \ref{fig:expansion-velocity} shows the distribution of the expansion velocities. The data points seem to concentrate around an expansion velocity 
of $\sim$8\kms.  

\subsection{The annual parallax distance and the secular motion of \i2248}
\label{sec:annual-parallax}

%%%%%%%%%%% Put here Tables 4 %%%%%%%%%%%%%
%%%%%%%%%%% Put here Figures 3 %%%%%%%%%%%%%

As mentioned in Section \ref{sec:spatio-kinematics}, in each of the observation epochs, the positions of all \h2o\ maser components (spots) in I22480 and that of J2254 were measured with respect to the position reference where the brightest maser spot in the map was located. Assuming that the position of J2254 has persisted as position reference, eventually the positions of the maser spots have been measured with respect to J2254.  We estimate the {\it intrinsic} position error of J2254 to be $\sigma_{\rm J2254}\approx$0.03--0.07~mas, which is statistically determined by the signal-to-noise ratio of detection of J2254, $\sigma\simeq 0.5\frac{\theta_{\rm beam}}{R_{\rm SN}}$. Here $\theta_{\rm beam}\simeq$1~mas is the synthesized beam size and $8\lesssim R_{\rm SN} \lesssim 20$ (for $S_{\nu}\approx$8--18~mJy) the signal-to-noise ratio \citep{mor93}. The {\it intrinsic} position error of the position-reference maser spot is much smaller than that of J2254 because of the much higher signal-to-noise ratio. The systematic error in each astrometric measurement has been discussed in previous papers dealing with VERA astrometry (e.g., \cite{hon10} and references therein). One of the major factors in the systematic error is attributed to uncertain zenith delay residuals of VERA stations caused by the atmosphere (contributions from the troposphere and the ionosphere, $\Delta\tau_{\rm atm}\approx\sqrt{\Delta\tau^{2}_{\rm tropo}+\Delta\tau^{2}_{\rm iono}}$), which are estimated to be  $\sigma_{\rm atm} \simeq \theta_{\rm beam}\frac{c\Delta\tau_{\rm atm}}{\lambda}\simeq 0.027 \theta_{\rm beam}\left(\frac{\theta_{\rm sep}}{\rm 1\arcdeg}\right)\approx $0.05~mas in the case of a typical zenith delay residual of about 2~cm \citep{hon08b} and a separation angle between the sources (1\arcdeg.94 between I22480 and J2254). Thus we estimate the position error of typically  $\sigma_{\rm 1epoch}\approx \sqrt{\sigma^{2}_{\rm atm}+\sigma^{2}_{\rm J2254}}\sim$0.08~mas in {\it each measurement}, but which was actually dependent on observation season. 

We attempted the least-square method for the spot motions to fit the modeled motions each of which is composed of an annual parallax, a constant secular motion, and an position offset at the reference epoch (J2000.0)\footnote
{In the model fitting, all spot positions were at first converted to the relative positions with respect to that at the first observation epoch for conserving the fitting accuracy.}. 
We used all of the epoch points except one without any valid astrometric result. Column 7 of table \ref{tab:status} shows valid (Y) and invalid (N) astrometric observation epochs. In the model fitting, weighting with position accuracy was adopted. The errors of maser spot positions and those for the kinematic model fitting parameters were adjusted by multiplying a constant factor to them so that a residual of the model fitting $\chi^{2}$ is reduced to unity. In order to obtain an annual parallax common in these spots, iterative procedures were adopted, which is similar to that adopted by \citet{san12}. At first, the model fitting was performed independently for the individual spot motions (independent fitting). At second, a mean annual parallactic motion was subtracted from the spot motions to estimate only the position offsets and secular motions of spots (proper-motion fitting). At third, the systemic motions estimated from the derived parameters were subtracted from the original spot motions. Finally, the position residuals were used to estimate a common parallax as well as position offset and proper motion residuals (combined fitting). The proper motion fitting was again performed using the estimated common parallax. Thus the procedures from the proper-motion to combined fitting were iterated until the estimated parameters were converged. The left and middle panels of Figure \ref{fig:parallax} show the motion of one of the target spots, the $-53.9$\kms\  component. The right panel of Figure \ref{fig:parallax} plots the positions of five maser spots with the individual linear proper motions subtracted. The modeled annual parallax modulation is superposed on the maser positions. Table \ref{tab:motion-model} gives the result of maser motion fitting.

Here we consider a possible systematic error caused by time variation of brightness strucutres of J2254 and the \h2o maser spots in I22480. We have found a simple, unresolved structure towards J2254, but which may be heavily affected by the synthesized beam pattern and low signal-to-noise ratio (Column 5 of Table \ref{tab:status}). In fact, the Gaussian brightness fitting errors for J2254 are larger than the intrinsic error mentioned above and they are highly variable and dependent on observation condition (see relative variation of error bar length in Figure \ref{fig:parallax}). This may be the largest contribution to the temporal fluctuation of the measured position (typically $\sigma_{\rm J2254}\sim$0.2~mas, up to 0.5~mas). The position of the position-reference maser spot is also fluctuated by temporal structure variation of the associating maser-emitting gas clump (feature).  As described in Appendix \ref{sec:appendix2}, we estimated such an error contribution to be typically $\sigma_{\rm maser}\approx$0.05~mas. Thus, we estimated a typical value of the spot position fluctuation (including a systematic position drift) to be $\sigma\approx\sqrt{\sigma^{2}_{\rm 1epoch}+\sigma^{2}_{\rm J2254}+\sigma^{2}_{\rm maser}}\sim$0.22~mas. 

In the motion fitting to each maser spot with a position flucturation amplitude of $\sim$0.22~mas in 10 data points (epochs), the accuracy of annual parallax achieves $\sigma_{\pi}\sim$0.05~mas, about 1/4 of the position errors of the individual data points. This is typically seen in previous VERA astrometry (e.g., \cite{and11}). Because some maser spots are associated with the same cluster of maser features, the position measurements for the five maser spots listed in Table \ref{tab:motion-model} are not completely independent. However, the fitting error contributed from the temporal (random) variation in maser feature structures (see Appendix \ref{sec:appendix2}) may be statistically mitigated. In fact, the final error of the determined annual parallax in the result of the combined fitting was reduced to $\sigma_{\pi}\approx$0.025~mas, smaller by a factor of about $\sqrt{4}$ than those in the indipendent fitting ($\sim$0.05~mas). Finally we obtained an annual parallax of $\pi=0.400\pm 0.025$~mas, corresponding to a distance to I22480 of $D=2.50^{+0.17}_{-0.15}$~kpc. 

We also estimated the secular motion of the star in I22480 to be $(\mu_{\alpha}, \mu_{\delta})=(-2.58\pm 0.33, -1.91\pm 0.17)$ [mas~yr$^{-1}$], which was derived from a secular motion of the $-53.9$\kms\ component, 
$(\mu_{\alpha}, \mu_{\delta})\equiv (\mu_{X}, \mu_{Y})=(-3.23\pm 0.05, -2.06\pm 0.05)$ [mas~yr$^{-1}$] (see Table 
\ref{tab:motion-model}), plus a relative motion of the star with respect to this maser spot $(\dot{X}, \dot{Y})=(0.65, 0.15)$ 
[mas~yr$^{-1}$] [corresponding to ($V_{{\rm 0}X}, V_{{\rm 0}Y}$)$=$(7, 2)[km~s$^{-1}$], see Section \ref
{sec:spatio-kinematics}]. The secular motion error is mainly attributed to the uncertainty of the stellar motion with respect to the maser spot motion. The westward proper motion of I22480 previously measured using several literatures \citep{ima08} is reconfirmed in this paper. 

\section{Discussion}
\label{sec:discussion}

As described in Sect. \ref{sec:spatio-kinematics}, the spatio-kinematical structure of \h2o masers in I22480 is well modeled by an expanding outflow with an expansion velocity of $\sim$7\kms. Such a flow is typically seen the circumstellar envelope around Mira variables (e.g., \cite{bow94}). On the other hand, the total extent of the maser feature distribution is up to $ \sim$170~AU ($D$/2.5~kpc), which is larger than those typically seen around Mira variables (up to 50~AU, e.g., \cite{bow94}) and comparable to or a little smaller than those seen around red supergiants (200--400~AU, e.g., \cite{cho08, asa10}). Therefore, these suggest that I22480 harbors an envelope as seen in M-type supergiants and supports the hypothesis of this star being a K-type supergiant, or a post-AGB star that still harbors its ejected envelope. I22480 was suggested to be an RV Tau-type variable star \citep{ima08}. However, the revised distance gives a revised value of the stellar luminosity of 35 000 $L_{\odot}$. Together with negative confirmation of optical magnitude variability, this suggests again that I22480 should be a K-type supergiant rather than an RV Tau variable. Note that \citet{win94} gave a new spectral classification of M0I for I22480. 

Obviously, the previous statistical parallax underestimated the distance to I22480 ($\sim$1~kpc, \cite{ima08}), 
in which only 13 proper motions were used for the estimation. Using the present data with 22 proper motions, this gives a distance value of 2.9$\pm$0.4~kpc, with rough agreement with the annual parallax distance (2.50~kpc). We have learnt that the statistical parallax method is valid only with a sufficient number of maser feature proper motions ($>$20), that are uniformly and independently distributed \citep{mor93}. 

The location and velocity vector of I22480 in the Milky Way are also estimated in the same manner as that adopted in \citet{ima11b} from the derived secular motion shown in Section \ref{sec:annual-parallax}. Table \ref{tab:MW-motion} gives the derived parameters. Although these values may change depending on the assumptions of the Galactic constants and the Solar motion with respect to the LSR, it is clear that the secular motion of I22480 is slower by $\sim$30\kms\ than a circular motion estimated from the galactocentric distance to I22480 and the assumption of a flat Galactic rotation curve, as commonly found in maser sources in the Perseus spiral arm \citep{rei09}. \citet{ima08} reported a larger peculiar motion, $U=-71$\kms\ and $V=-29$\kms\ for I22480, where $U$ and $V$ indicate the peculiar motion components in the anti-galactic center and galactic rotation directions, respectively. However, these values also should be revised to those determined in this paper ($U=-11$\kms\ and $V=-22$\kms) because the distance and the secular motion of I22480 are now revised from 1.0~kpc to 2.5~kpc and from $\sim$17~mas~yr$^{-1}$ in the west direction to $\sim$3~mas~yr$^{-1}$ in the southwest direction, respectively. In addition, I22480 is located close to the Galactic mid-plane ($z\sim$60~pc). They also support the hypothesis of a (high-mass) K-type supergiant in I22480 being a member of the Galactic thin disk. The discussion of the Galactic rotation curve itself based on the peculiar motions of \h2o\ maser sources including that of I22480 will be described in a separate paper. 

\bigskip
VERA/Mizusawa VLBI observatory is a branch of the National Astronomical Observatory of Japan, an interuniversity research institute operated by the Ministry of Education, Culture, Sports, Science and Technology. 
We acknowledge all staff members and students who have helped in array operation and in data correlation of the VERA. H.~I. was supported by Grant-in-Aid for Scientific Research from the Ministry of Education, Culture, Sports, Science, and Technology (18740109) and the Strategic Young Researcher Overseas Visits Program for Accelerating Brain Circulation funded by Japan Society for the Promotion of Science. 

\appendix
\section{Inverse phase-referencing method in VERA}
\label{sec:appendix1}

As described in Section \ref{sec:observations}, the present work adopted the inverse phase-referencing (IPR) technique for the first time throughout the whole observation epochs. This technique employs the transfer of fringe phase (and rate) calibration solutions from maser to continuum source data instead of that from the continuum to maser source data employed in normal phase-referencing (PR) technique for maser source astrometry. The IPR technique has been often adopted in maser source astrometry with the Very Long Baseline Array (VLBA)(e.g., \cite{rei09,sat10}). In the VLBA observation, a maser and a position-reference continuum sources are observed in the same signal-processing path and in the same frequency setup. In this case, the data from another scans on group-delay calibrators can commonly used for calibrating both source data. Fringe fitting with the data of the maser scan is performed after this group-delay calibration in order to obtain only phase and rate residuals. These solutions are applied to the data of the position-reference source. 

\begin{figure*}[p]
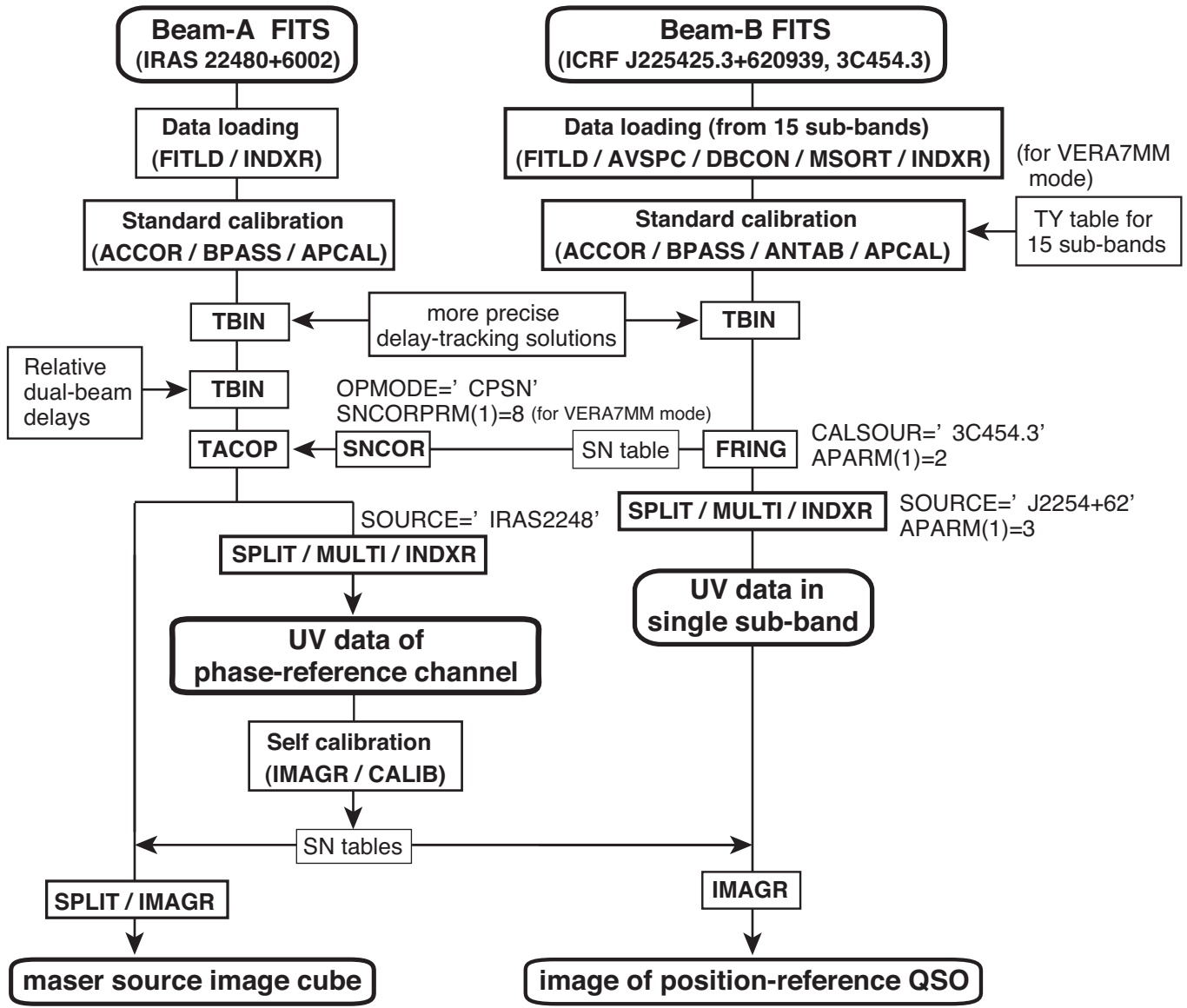

  \begin{center}
       \FigureFile(\hsize,\hsize){fig5.eps}
  \end{center}
  \caption{Flow chart of the data reduction and image synthesis in the case of the {\it inverse phase-referencing technique} for VERA data. A round-square block denotes a visibility (UV) or image cube data set. A thick-black square block denotes a segment of process in AIPS. Setting of the essential parameters (adverbs) is described on the side. A thin-black square block denotes a file or solution (SN) table loaded or used for data calibration. The parts of calibration solution application (CLCAL in AIPS) are just expressed by crossing points of the lines that show the flows of data processing.}
\label{fig:analysis}
\end{figure*}

In the case of VERA, not only different signal processing paths but also different frequency setup are used between maser and continuum signals. Therefore, a sequence of data calibration steps is different from that in the VLBA data analysis. Figure \ref{fig:analysis} briefly shows the sequence of the whole data reduction in the IPR technique for VERA data. The essential and unique parts of the VERA IPR technique are summarized as follows.

\begin{itemize}
\item The solutions for more precise delay tracking and those of the relative instrumental delays between Beam-A and -B should be applied before any delay and phase calibration using the visibility data themselves. 
\item The group-delay calibration shall be made by using the data of another continuum calibrator, which should be brighter than the position-reference continuum source. The calibrator signals shall be recorded in Beam-B (15 BBCs), which have a much wider total bandwidth than Beam-A does (only one BBC). The group-delay calibration solutions shall be obtained by calculating multi-BBC group delay residuals. The useful phase calibration solutions shall be contained in the eighth BBC (in the digital filter mode of VERA7MM) of the solution (SN) table, whose base frequency is coincident with that in Beam-A, and should be copied to the first sub-band. When they are copied to the data set of Beam-A, only those in the first sub-band are recognized in Beam-A. 
\item The solutions of fringe-fitting and self-calibration using bright maser emission in Beam-A shall be applied to the data set of Beam-B {\it after} the Beam-B data are integrated into a single BBC. 
\end{itemize}

The ParselTongue/python scripts for the PR/IPR calibration for VERA data in AIPS are now available in the wiki page\footnote
{http://milkyway.sci.kagoshima-u.ac.jp/groups/vcon\_lib/wiki/9fbfd/Data\_Analysis.html}
\footnote
{See also the ParselTongue wiki page: http://www.jive.nl/dokuwiki/doku.php?id=parseltongue:parseltongue}.

\section{Contribution of maser structures to the accuracy of absolute astrometry}
\label{sec:appendix2}
\begin{figure*}[p]
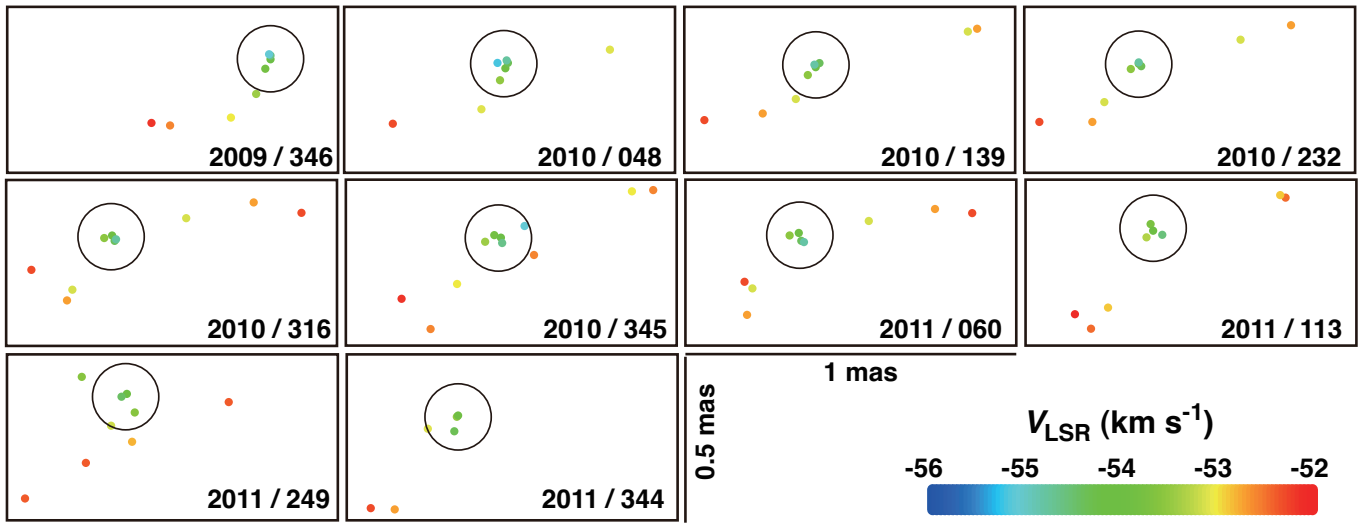

  \begin{center}
       \FigureFile(\hsize,\hsize){fig6.eps}
  \end{center}
  \caption{Maser spot distribution around the $-53.9$--$-54.1$\kms\ component in the maser feature 
  \i2248:I2013-{\it 4}, which was one of the maser spots used for the annual parallax measurement, 
  at all observation epochs within a box of 1~mas $\times$ 0.5~mas. A opened circle with a radius 
  of 0.1~mas is centered at the position-reference spot. 
  The date of observation epoch is shown in the bottom-right corner of each sub-panel. } 
\label{fig:reference-spot}
\end{figure*}

\h2o\ maser spots or features, as reference candles for VLBI astrometry, are time-variable. It is impossible to visualize the whole maser feature structure in VLBI imaging and the locations of the individual maser spots in the feature because we are always watching only the brightest parts of the feature structure in the spectral channel maps. Nevertheless, we can evaluate the stability of the structure and the accuracy of the astrometry described in this paper. Figure \ref{fig:reference-spot} shows the relative distribution of maser spots including one of the position-reference maser spots at all epochs. The position-reference spot was always located around the center of the whole distribution of maser spots (or a maser feature) whose spatial and velocity pattern has persisted within $\sim$0.1~mas. In particular, the locations of the position-reference spot and its velocity-adjacent spots (within 0.4\kms) were always concentrated within $\sim$0.05~mas, corresponding to 0.6\kms\ at 2.5~kpc from the Sun, although other spots exhibit relatively rapid variation in their locations. The latter scale may give the accuracy of our astrometry. Because the velocity spacing of VLBI spectroscopy itself was about 0.4\kms (Section \ref{sec:observations}), and the corresponding spectral channels covered slightly different velocity ranges among the observation epochs, we consider a group of maser spots within 1\kms\ in our astrometry. 

\begin{figure*}[p]
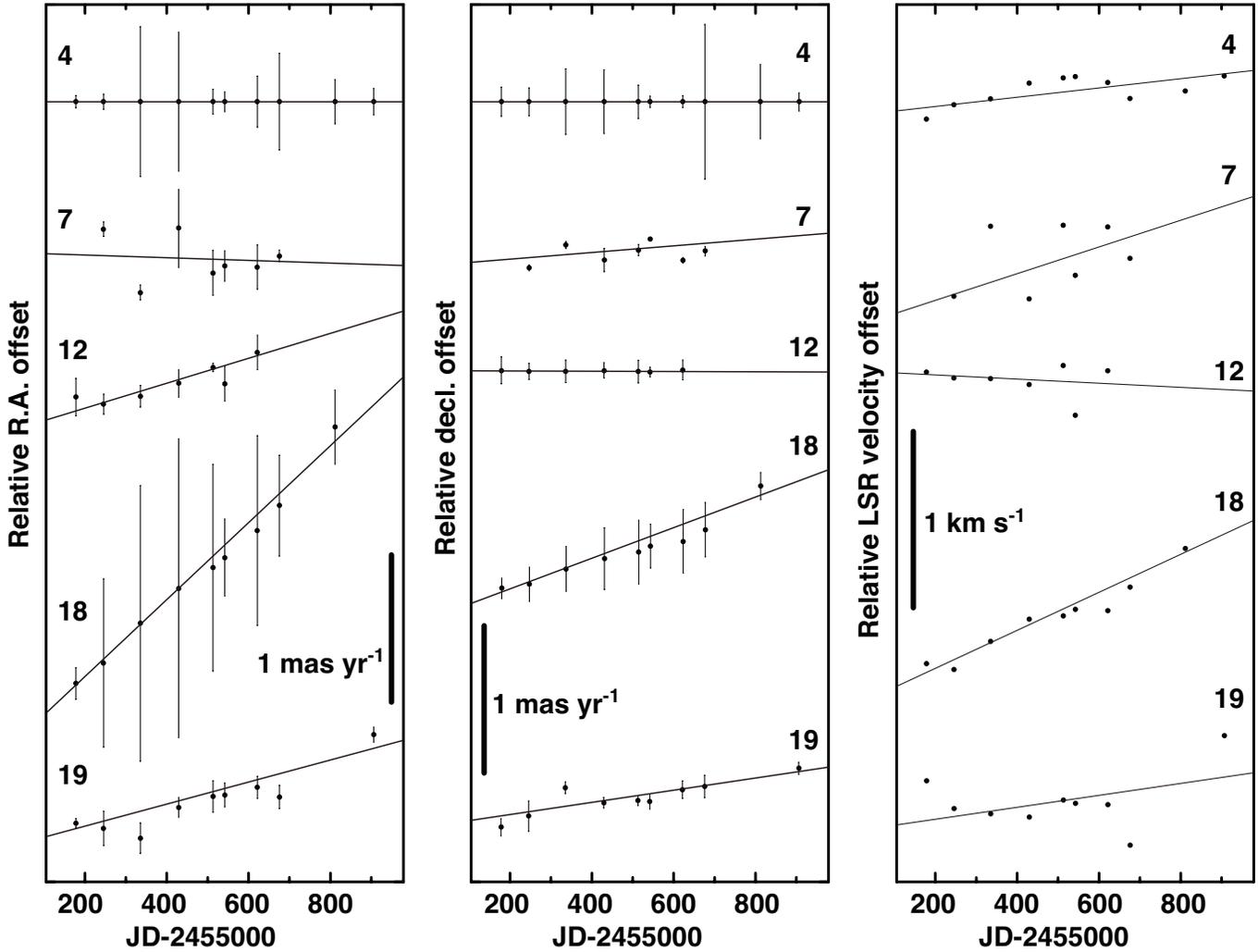

  \begin{center}
       \FigureFile(\hsize,\hsize){fig7.eps}
  \end{center}
  \caption{Observed relative proper motions and LSR velocity drifts of the \h2o maser features in \i2248, 
  whose accompanying brightest maser spots were the targets of the annual parallax measurement. 
  The number added for each proper motion shows the assigned one after the designated name form 
  g\i2248:I2013h. A thin solid line indicates a least-squares-?tted line assuming a constant velocity 
  proper motion and a constant rate of the LSR velocity drift. A vertical bar for each data point in the left 
  and middle panels indicates the defined size of the maser feature. For clarity, the velocity widths of 
  maser features are not displayed.} 
\label{fig:feature-motions}
\end{figure*}

This is also supported by the deviations of maser feature positions from fitted ballistic motions (or constant proper motions) of the features. Figure \ref{fig:feature-motions} shows such deviations for the maser features whose accompanying brightest maser spots were the targets of the annual parallax measurement. The proper motions were measured with respect to the reference feature \i2248:I2013-{\it 4} (with zero relative proper motion). The displayed position and size of maser feature were derived as shown in \citet{ima02}. Note that we suppose constant rates of LSR velocity drifts of maser features. Even the largest drift rate $\sim$0.4\kms~yr$^{-1}$, however, corresponds to only 0.03~mas~yr$^{-2}$ in rate of proper motion variation per year. Except a small number of large deviation cases (up to 0.3~mas), the position deviations from the expected ballistic feature motions are also within $\sim$0.05~mas.  The origin of the position deviation and the LSR velocity drifts are expected to be {\it micro-turbulence} in the features \citep{ima02}.  

%\clearpage

%%%%%%%%%%% Table 1 %%%%%%%%%%%%%%%%%%%
\begin{table*}[ht]
\caption{Parameters of the VERA observations.}\label{tab:status}
\scriptsize
\begin{tabular}{l@{ }l@{ }c@{ }c@{ }l@{ }rc@{ }c@{ }r@{ }r@{ }l} \hline \hline
& & & & & & \multicolumn{5}{c}{Astrometry} \\
& & & & & & \multicolumn{5}{c}{\ \hrulefill} \\
\multicolumn{2}{c}{Observation} & VERA & & \multicolumn{1}{c}{Beam\footnotemark[c]} 
&  &  & $V_{\rm ref}$\footnotemark[f] & $F_{\rm ref}$\footnotemark[g] & $I_{\rm J2254}$\footnotemark[h] &\\
code & epoch  & telescopes\footnotemark[a] & Noise\footnotemark[b]
&  \multicolumn{1}{c}{[mas]} & $N_{\mbox{\scriptsize s}}$\footnotemark[d] 
& valid?\footnotemark[e] & [km~s$^{-1}$] & [Jy] & [mJy/b] & Note \\ \hline 
r09346a \dotfill & 2009 December 12 & MROS & 30
& 1.18$\times$0.93, $-50^{\circ}$\hspace{-2pt}.5 & 19 & Y & $-$53.8 & 20 & 10.3 & \\
r10048b \dotfill & 2010 February 18 & MROS & 30   
& 1.20$\times$0.86, $-45^{\circ}$\hspace{-2pt}.8 & 14 & Y & $-$53.9 & 24 & 8.3 & \\
r10139a \dotfill & 2010 May 19 & MROS & 107
& 1.22$\times$0.80, $-28^{\circ}$\hspace{-2pt}.9 & 9 & Y & $-$53.9 & 36 & 13.6 & \\ 
r10232a \dotfill & 2010 August 10 & MROS & 80
& 1.19$\times$0.93, $-31^{\circ}$\hspace{-2pt}.6 & 9 & Y & $-$53.9 & 24 & 13.5 & \\ 
r10316b \dotfill & 2010 November 12 & MROS &  64
& 1.51$\times$0.91, $-45^{\circ}$\hspace{-2pt}.9 & 13 & Y & $-$53.9 & 22 & 17.9 & \\ 
r10345a \dotfill & 2010 December 11 & MROS & 44
& 1.23$\times$0.89, $-44^{\circ}$\hspace{-2pt}.5 & 17 & Y& $-$53.8 & 19 & 16.1 & \\ 
r11060c \dotfill & 2011 March 2 & MROS & 48
& 1.42$\times$0.86, $-46^{\circ}$\hspace{-2pt}.6 & 20 & Y & $-$53.9 & 12 & 17.5 & \\ 
r11113b \dotfill & 2011 April 23 & MROS &  46
& 1.38$\times$0.92, $-28^{\circ}$\hspace{-2pt}.1 & 7 & Y & $-$54.1 & 6 & 16.0 & \\ 
r11215a \dotfill & 2011 August 3 & (M)ROS & --- 
& \multicolumn{2}{c}{---} & N & --- & --- & & Bad recording in Mizusawa. \\ 
r11249b \dotfill & 2011 September 6 & MROS & 80 
& 1.42$\times$0.93, $-62^{\circ}$\hspace{-2pt}.4 & 7 & Y & $-$51.0 & 6 & 13.7  & \\ 
r11344a \dotfill & 2011 December 10 & MROS & 47
& 1.23$\times$1.05, $-55^{\circ}$\hspace{-2pt}.1 & 11 & Y & $-$51.4 & 7 & 13.7 & \\ 
\hline
\end{tabular}

\footnotemark[a]Telescope whose data were valid for phase-referencing maser imaging. 
M: Mizusawa, R: Iriki, O: Ogasawara, S: Ishigakijima.
The station with parentheses had some problem during the observations and affected the annual parallax measurements. \\
\footnotemark[b]rms noise  in the emission-free spectral channel image in units of mJy beam$^{-1}$.\\ 
\footnotemark[c]Synthesized beam size resulting from natural weighted visibilities, i.e. major and minor axis lengths and position angle.\\
\footnotemark[d]Number of the identified maser features.\\
\footnotemark[e]Y and N: valid and invalid data point for annual parallax measurement, respectively. \\
\footnotemark[f]LSR velocity of the spectral channel including the maser spot used for fringe-fitting, 
self-calibration, and the inverse phase referencing (see the main text). \\
\footnotemark[g]Flux density of the maser spot used for fringe-fitting, 
self-calibration, and the inverse phase referencing. \\
\footnotemark[h]Peak intensity of \j2254 in units of mJy~beam$^{-1}$ on the phase-referenced image. \\
\end{table*}

%%%%%%%%%%% Table 2 %%%%%%%%%%%%%%%%%%%
\begin{table*}[h]
\caption{Parameters of the \h2o  maser features identified by 
proper motion toward \i2248} \label{tab:pmotions}
\scriptsize
\begin{tabular}{lrrrrrrrr    c@{ }c@{ }c@{ }c@{ }c@{ }c@{ }c@{ }c@{ }c@{ }c } \hline \hline          
 & \multicolumn{2}{c}{Offset}
 & \multicolumn{4}{c}{Proper motion\footnotemark[b]}
 & \multicolumn{2}{c}{Radial motion\footnotemark[c]}
 & \\                                                                                                
 & \multicolumn{2}{c}{(mas)} 
 & \multicolumn{4}{c}{(mas yr$^{-1}$)}
 & \multicolumn{2}{c}{(km s$^{-1}$)}
 & \multicolumn{10}{c}{Detection at 10 epochs} \\                                               
 & \multicolumn{2}{c}{\ \hrulefill \ } 
 & \multicolumn{4}{c}{\ \hrulefill \ } 
 & \multicolumn{2}{c}{\ \hrulefill \ } 
 & \multicolumn{10}{c}{\ \hrulefill \ } \\                                                           
 Feature\footnotemark[a] & $\Delta$R.A. & $\Delta$decl. & $\mu_{x}$ & $\sigma \mu_{x}$ & $\mu_{y}$ 
 & $\sigma \mu_{y}$ & V$_{z}$ & $\Delta$V$_{z}$\footnotemark[d] 
 & 1& 2& 3& 4& 5& 6& 7& 8& 9&10 \\ \hline                                                          
  1   \ \dotfill \  &$    29.78$&$     4.05$&$   2.22$&   0.15 &$  -0.96$&   0.24
 &$ -58.42$&   2.11
 &$\circ$ &$\circ$ &$\circ$ &$\circ$ &$\times$ &$\times$ &$\times$ &$\times$ &$\times$ &$\times$   \\
  2   \ \dotfill \  &$    31.46$&$     3.41$&$   1.00$&   0.41 &$  -0.20$&   0.37
 &$ -55.93$&   2.00
 &$\times$ &$\times$ &$\times$ &$\times$ &$\circ$ &$\circ$ &$\circ$ &$\circ$ &$\times$ &$\times$   \\
  3   \ \dotfill \  &$    -1.42$&$     3.02$&$   0.19$&   0.29 &$   0.36$&   0.22
 &$ -55.05$&   0.70
 &$\times$ &$\times$ &$\times$ &$\times$ &$\times$ &$\circ$ &$\circ$ &$\circ$ &$\times$ &$\times$   \\
  4   \ \dotfill \  &$     0.00$&$     0.00$&$   0.00$&   0.04 &$   0.00$&   0.05
 &$ -54.04$&   2.95
 &$\circ$ &$\circ$ &$\circ$ &$\circ$ &$\circ$ &$\circ$ &$\circ$ &$\circ$ &$\circ$ &$\circ$   \\      
  5   \ \dotfill \  &$    -0.42$&$     2.56$&$  -0.12$&   0.42 &$  -0.50$&   0.27
 &$ -53.36$&   0.21
 &$\circ$ &$\circ$ &$\times$ &$\times$ &$\times$ &$\times$ &$\times$ &$\times$ &$\times$ &$\times$  \\ 
 
  6   \ \dotfill \  &$    -2.21$&$    -2.01$&$  -0.04$&   0.27 &$  -0.27$&   0.16
 &$ -53.08$&   0.79
 &$\times$ &$\times$ &$\times$ &$\circ$ &$\circ$ &$\circ$ &$\circ$ &$\times$ &$\times$ &$\times$   \\
  7   \ \dotfill \  &$    -0.79$&$     0.13$&$  -0.03$&   0.05 &$   0.08$&   0.02
 &$ -53.06$&   0.75
 &$\times$ &$\circ$ &$\circ$ &$\circ$ &$\circ$ &$\circ$ &$\circ$ &$\circ$ &$\times$ &$\times$   \\   

 8   \ \dotfill \  &$     0.28$&$    -0.37$&$   0.24$&   1.08 &$  -0.21$&   1.17
 &$ -52.94$&   1.48
 &$\times$ &$\times$ &$\times$ &$\times$ &$\times$ &$\circ$ &$\circ$ &$\times$ &$\times$ &$\times$   \\
 9   \ \dotfill \  &$     0.23$&$     1.95$&$   1.76$&   0.31 &$   0.80$&   0.12
 &$ -52.10$&   0.98
 &$\circ$ &$\circ$ &$\circ$ &$\times$ &$\times$ &$\times$ &$\times$ &$\times$ &$\times$ &$\times$   \\
 10   \ \dotfill \  &$     0.78$&$    -0.44$&$   0.37$&   0.13 &$   0.03$&   0.02
 &$ -51.68$&   1.69
 &$\circ$ &$\circ$ &$\times$ &$\times$ &$\circ$ &$\circ$ &$\circ$ &$\times$ &$\times$ &$\times$   \\ 
 11   \ \dotfill \  &$     0.67$&$    -0.58$&$  -0.06$&   0.42 &$  -0.15$&   0.58
 &$ -51.20$&   2.53
 &$\times$ &$\times$ &$\times$ &$\times$ &$\times$ &$\times$ &$\times$ &$\times$ &$\circ$ &$\circ$   \\
 12   \ \dotfill \  &$     0.01$&$    -0.73$&$   0.31$&   0.08 &$   0.00$&   0.06
 &$ -51.01$&   1.93
 &$\circ$ &$\circ$ &$\circ$ &$\circ$ &$\circ$ &$\circ$ &$\circ$ &$\circ$ &$\times$ &$\times$   \\   
 13   \ \dotfill \  &$    28.40$&$     5.45$&$   1.24$&   0.27 &$   0.83$&   0.22
 &$ -50.83$&   0.49
 &$\times$ &$\times$ &$\times$ &$\times$ &$\times$ &$\circ$ &$\circ$ &$\circ$ &$\times$ &$\times$   \\
 14   \ \dotfill \  &$    32.28$&$     5.20$&$   0.92$&   0.23 &$  -0.12$&   0.18
 &$ -50.55$&   0.42
 &$\times$ &$\times$ &$\times$ &$\times$ &$\circ$ &$\circ$ &$\circ$ &$\times$ &$\times$ &$\times$   \\
 15   \ \dotfill \  &$    29.74$&$    -6.19$&$   1.29$&   0.23 &$  -1.36$&   0.26
 &$ -49.69$&   0.21
 &$\times$ &$\circ$ &$\circ$ &$\times$ &$\times$ &$\times$ &$\times$ &$\times$ &$\times$ &$\times$  \\ 
 16   \ \dotfill \  &$    29.34$&$     3.65$&$   0.22$&   0.46 &$   0.10$&   0.12
 &$ -49.57$&   0.84
 &$\circ$ &$\times$ &$\times$ &$\circ$ &$\times$ &$\times$ &$\times$ &$\times$ &$\times$ &$\times$   \\
 17   \ \dotfill \  &$    30.38$&$     6.39$&$  -0.22$&   1.44 &$  -0.12$&   0.56
 &$ -49.21$&   1.68
 &$\times$ &$\times$ &$\times$ &$\times$ &$\times$ &$\times$ &$\times$ &$\times$ &$\circ$ &$\circ$   \\
 18   \ \dotfill \  &$    29.48$&$     6.21$&$   0.96$&   0.13 &$   0.38$&   0.06
 &$ -49.17$&   2.39
 &$\circ$ &$\circ$ &$\circ$ &$\circ$ &$\circ$ &$\circ$ &$\circ$ &$\circ$ &$\circ$ &$\times$   \\     
 19   \ \dotfill \  &$     4.77$&$     8.05$&$   0.28$&   0.03 &$   0.15$&   0.03
 &$ -47.00$&   1.12
 &$\circ$ &$\circ$ &$\circ$ &$\circ$ &$\circ$ &$\circ$ &$\circ$ &$\circ$ &$\times$ &$\circ$   \\     
 20   \ \dotfill \  &$     4.86$&$    13.43$&$  -0.08$&   1.41 &$  -0.07$&   0.68
 &$ -44.94$&   0.84
 &$\times$ &$\times$ &$\times$ &$\times$ &$\times$ &$\circ$ &$\circ$ &$\times$ &$\times$ &$\times$   \\
 21   \ \dotfill \  &$    29.55$&$    11.67$&$   0.38$&   0.33 &$   0.06$&   0.27
 &$ -44.51$&   0.31
 &$\times$ &$\times$ &$\times$ &$\times$ &$\times$ &$\circ$ &$\circ$ &$\times$ &$\times$ &$\times$   \\
 22   \ \dotfill \  &$    39.41$&$   -47.15$&$   1.12$&   0.33 &$  -0.77$&   0.43
 &$ -43.67$&   1.05
 &$\circ$ &$\circ$ &$\times$ &$\times$ &$\times$ &$\times$ &$\times$ &$\times$ &$\times$ &$\times$   \\
 \hline
 \end{tabular}

\footnotemark[a]\h2o maser features detected toward \i2248. The feature is designated as \i2248:I2013-{\it N}, 
where {\it N} is the ordinal source number given in this column (I2013 stands for sources found by Imai 
\etal\ and listed in 2013).\\
\footnotemark[b]Relative value with respect to the position-reference maser feature: \i2248:I2013-{\it 4}. \\
\footnotemark[c]Relative value with respect to the local standard of rest. \\
\footnotemark[d]Mean full velocity width of amaser feature at half intensity. \\
\end{table*}

%\clearpage
%%%%%%%%%%% Table 3 %%%%%%%%%%%%%%%%%%%
\begin{table*}[h]
\caption{Parameters of the best fit 3D spatio-kinematical model of the \h2o\ masers in \i2248}
\label{tab:kinematic-model}

\scriptsize
\begin{tabular}{lr@{$\pm$}l} \hline \hline
Parameter \hspace{1.5cm} & \multicolumn{2}{c}{Value} \\ \hline
$\Delta X_{0}$ [mas] \footnotemark[a] \dotfill & 22 & 8 \\
$\Delta Y_{0}$ [mas] \footnotemark[a] \dotfill & 9 & 6 \\
$V_{0X}$ [km~s$^{-1}$] \footnotemark[a] \dotfill & 7 & 4 \\
$V_{0Y}$ [km~s$^{-1}$] \footnotemark[a] \dotfill & 2 & 2 \\
$V_{0Z}$ [km~s$^{-1}$] \footnotemark[b] \dotfill & \multicolumn{2}{c}{$-50$} \\ 
$D$ [kpc] \footnotemark[b] \  \dotfill & \multicolumn{2}{c}{2.5} \\ \hline
$\sqrt{\chi^{2}}$ \footnotemark[c] \dotfill & \multicolumn{2}{c}{2.08}  \\ \hline
\end{tabular}

\footnotemark[a]Relative value respect to the maser feature \i2248:I2013-{\it 4}. \\
\footnotemark[b]Assumed value. \\
\footnotemark[c]The $\chi^{2}$-square value was obtained in the model fitting based on the least-squares method. \\
%\end{table*}

%%%%%%%%%%% Table 4 %%%%%%%%%%%%%%%%%%%
%\begin{table*}[h]
\caption{Parameters of the fitted maser spot motion}
\label{tab:motion-model}

\scriptsize
\begin{tabular}{lr@{$\pm$}l r@{$\pm$}l r@{$\pm$}l r@{$\pm$}l r@{$\pm$}lrr} \hline \hline
\multicolumn{1}{c}{\vlsr} & \multicolumn{2}{c}{$X_{\mbox \scriptsize 0}$\footnotemark[a]} 
& \multicolumn{2}{c}{$Y_{\mbox \scriptsize 0}$\footnotemark[a]}
& \multicolumn{2}{c}{$\mu_{X}$} & \multicolumn{2}{c}{$\mu_{Y}$} & \multicolumn{2}{c}{$\pi$} 
& \multicolumn{2}{c}{Deviation\footnotemark[b]} \\
\multicolumn{1}{c}{[km~s$^{-1}$]} & \multicolumn{2}{c}{[mas]} 
& \multicolumn{2}{c}{[mas]} & \multicolumn{2}{c}{[mas~yr$^{-1}$]}
& \multicolumn{2}{c}{[mas~yr$^{-1}$]} & \multicolumn{2}{c}{[mas]}  & $\sigma_{X}$ & $\sigma_{Y}$ \\ \hline
\multicolumn{11}{c}{Independent fitting} \\ \hline
$-53.9$ & 32.48 &    0.56 &   20.33 &    0.57 & $  -3.23$ &   0.05 & $  -2.04$ &  0.05 &  0.400 &  0.040 & 0.13 & 0.11 \\
$-52.7$ & 32.78 &    1.72 &   19.99 &    1.99 & $  -3.23$ &   0.16 & $  -1.95$ &   0.18 &  0.398 &  0.071 & 0.25 & 0.17 \\
$-51.0$ & 28.51 &    1.01 &   22.28 &    1.11 & $  -2.84$ &   0.10 & $  -2.23$ &  0.10 &  0.428 &  0.059 & 0.17 & 0.09 \\
$-48.9$ & 22.32 &    0.69 &   18.76 &    0.72 & $  -2.22$ &   0.06 & $  -1.87$ &   0.07 &  0.380 &  0.038 & 0.12 & 0.08 \\
$-47.2$ & 29.06 &    0.94 &   20.75 &    0.99 & $  -2.90$ &   0.09 & $  -2.01$ &  0.09 &  0.316 &  0.052 & 0.14 & 0.14 \\
\hline
\multicolumn{11}{c}{Proper motion fitting (after subtracting the annual parallax\footnotemark[c]} \\ \hline
 $-53.9$ & 32.63 &    1.45 &   20.71 &    1.30 & $  -3.23$ &   0.13 & $  -2.06$ &  0.12 &  
 \multicolumn{2}{c}{...} & \multicolumn{2}{c}{...}\\
 $-52.7$ & 34.00 &    3.45 &   20.92 &    2.72 & $  -3.33$ &   0.32 & $  -2.03$ &   0.25 &
 \multicolumn{2}{c}{...} & \multicolumn{2}{c}{...}\\
 $-51.0$ & 30.42 &    1.85 &   21.54 &    2.13 & $  -3.01$ &   0.17 & $  -2.15$ &  0.20 &  
 \multicolumn{2}{c}{...} & \multicolumn{2}{c}{...}\\
 $-48.9$ & 24.47 &    2.09 &   17.93 &    1.91 & $  -2.40$ &   0.19 & $  -1.78$ &  0.18 &  
 \multicolumn{2}{c}{...} & \multicolumn{2}{c}{...}\\
 $-47.2$ & 31.54 &    2.48 &   19.12 &    2.23 & $  -3.11$ &   0.23 & $  -1.90$ &  0.21 &  
 \multicolumn{2}{c}{...} & \multicolumn{2}{c}{...}\\
\hline
\multicolumn{11}{c}{Combined fitting to an annual parallax} \\ \hline
Combined & $-1.19$ & 0.43 & $-$0.07 & 0.45 & $  0.10 $ &  0.04 & $-0.02 $ &  0.04 &   0.400 & 0.025 & 0.18 & 0.13 \\
\hline
\end{tabular}

\footnotemark[a]Position at the epoch J2000.0 with respect to the delay-tracking center (see main text).\\
\footnotemark[b]Mean deviation of the data points from the model-fit motion in unit of mas. \\
\footnotemark[c]The adopted annual parallax was given in the combined fitting in the converging phase of iterations. \\
%\end{table*}

%%%%%%%%%%% Table 5 %%%%%%%%%%%%%%%%%%%
%\begin{table*}[h]
\caption{Location and 3D motion of \i2248 in the Milky Way estimated from the VERA astrometry}
\label{tab:MW-motion}

\scriptsize
\begin{tabular}{lr@{$\pm$}l@{}} \hline \hline
Parameter \hspace{3.8cm} & \multicolumn{2}{c}{Value} \\ \hline
Galactic coordinates, $(l,b)$ [deg]\footnotemark[a] \dotfill  & \multicolumn{2}{c}{(108.43, 0.89)} \\
Heliocentric distance, $D$ [kpc]\footnotemark[a] \dotfill  & 2.50 & 0.16 \\
Systemic LSR velocity, $V_{\rm sys}$ [km~s$^{-1}$]\footnotemark[a] \dotfill & $-50.8$ & 3.5 \\
Secular proper motion, $\mu_{\rm RA}$ [mas~yr$^{-1}$] \dotfill & $-2.58$ & 0.33 \\
\hspace*{30mm} $\mu_{\rm decl}$ [mas~yr$^{-1}$] \dotfill & $-1.91$ & 0.17 \\
$R_{\rm 0}$ [kpc]\footnotemark[b] \dotfill  & \multicolumn{2}{c}{8.0} \\ 
$\Theta_{\rm 0}$ [km~s$^{-1}$]\footnotemark[b] \dotfill & \multicolumn{2}{c}{220} \\
$(U_{\odot}, V_{\odot}, W_{\odot})$ [km~s$^{-1}$]\footnotemark[c] \dotfill 
& \multicolumn{2}{c}{(7.5, 13.5, 6.8)} \\
$z_{\rm 0}$ [pc]\footnotemark[d] \dotfill & \multicolumn{2}{c}{16} \\
$R_{\rm gal}$ [kpc] \dotfill & 9.10 & 0.09 \\
$z$ [pc] \dotfill & 55 & 2 \\
$V_{R}$ [km~s$^{-1}$]  \dotfill & $-$11 & 4 \\
$V_{\theta}$ [km~s$^{-1}$]  \dotfill & 197 & 4 \\
$V_{z}$ [km~s$^{-1}$]  \dotfill & 0 & 3 \\ \hline
\end{tabular}

\footnotemark[a]Input value for \i2248.\\
\footnotemark[b]Input value for the Sun in the Milky Way. \\
\footnotemark[c]Motion of the Sun with respect to the local standard of rest, 
cited from \citet{fra09} (c.f., \cite{deh98}). \\
\footnotemark[d]Height of the Sun from the Galactic mid-plane, cited from \citet{ham95}. \\
\end{table*} 

\clearpage
%%%%%%%%%%%%%%%% Figures %%%%%%%%%%%%%%%%%
%%%%%%%% Figure 1 %%%%%%%%%%%%%%%%%

%%%%%%%% Figure 2 %%%%%%%%%%%%%%%%%

%%%%%%%% Figure 3 %%%%%%%%%%%%%%%%%

%%%%%%%% Figure 4 %%%%%%%%%%%%%%%%%
\begin{figure*}[p]
  \begin{center}
       \FigureFile(\hsize,\hsize){fig4.eps}
  \end{center}
  \caption{(a) Motion of the $-53.9$\kms\ component of \h2o maser emission in \i2248 and the kinematical model fit to this motion. R.A. and decl. offsets with respect to the phase-tracking center of the $-53.9$\kms\ component. A filled circle shows the data point observed and used for the annual parallax measurement. A solid curve shows the modeled motion including an annual parallax and a constant velocity proper motion. {\bf The length of horizontal and vertical bars in each data point indicates the uncertainty of spot position, which is adjusted so that a $\chi^2$ value in the model fitting would be reduced to unity.} An opened circle indicates the spot position expected in the model at the observation epoch. (b) Same as (a) but right ascension and declination offset variation of the maser spot along time. The estimated linear proper motion is subtracted from the observed spot position. A solid curve shows the modeled annual parallactic motion. (c) The result of the combined annual parallax fitting using five maser spots at \vlsr$=$ $-53.9$, $-52.7$, $-51.0$, $-48.9$, and $-47.2$\kms\ in the maser features \i2248:I2013-{\it 4}, {\it 7}, {\it 12}, {\it 18}, and {\it 19}, respectively. For clarity, ata points in different spots are denoted in different colors and slightly shifted in the horizontal axis.}
\label{fig:parallax}
\end{figure*}

%%%%%%%% Figure 5 %%%%%%%%%%%%%%%%%

\clearpage 
%\setlength{\baselineskip}{6ex}

%%%%%%%% Figure 6 %%%%%%%%%%%%%%%%%

\clearpage
%%%%%%%% Figure 7 %%%%%%%%%%%%%%%%%

\end{document}